\documentstyle[11pt,aaspp4,psfig]{article}    

\begin{document}

\title{X-ray Images of Hot Accretion Flows}

\author{Feryal \"Ozel\altaffilmark{1} \& Tiziana Di Matteo\altaffilmark{2}}
\affil{Harvard-Smithsonian Center for Astrophysics, 60 Garden 
Street, Cambridge, MA 02138; \\ fozel, tdimatteo@cfa.harvard.edu}
\altaffiltext{1}{Physics Department, Harvard University}
\altaffiltext{2}{{\em Chandra} Fellow}

\begin{abstract}
  
  We consider the X-ray emission due to bremsstrahlung processes from
  spherically-symmetric low radiative-efficiency hot accretion flows
  around supermassive and galactic black holes. We calculate surface
  brightness profiles and Michelson visibility functions for a range
  of density profiles, $\rho \sim r^{-3/2+p}$, with $0 < p < 1$, to
  allow for the presence of outflows. We find that although the 1~keV
  emitting region in these flows can always extend up to $10^6$
  Schwarzschild radii ($R_{\rm{S}}$), their surface brightness
  profiles and visibility functions are strongly affected by the
  specific density profile. The advection-dominated solutions with no
  outflows ($p=0$) lead to centrally peaked profiles with
  characteristic sizes of only a few tens of $R_{\rm{S}}$. Solutions
  with strong outflows ($p \sim 1$) lead to flat intensity profiles
  with significantly larger characteristic sizes of up to $10^6
  R_{\rm{S}}$. This implies that low luminosity galactic nuclei, such
  as M87, may appear as extended X-ray sources when observed with
  current X-ray imaging instruments.  We show that X-ray brightness
  profiles and their associated visibility functions may be powerful
  probes for determining the relevant mode of accretion and, in turn,
  the properties of hot accretion flows. We discuss the implications
  of our results for observations with the {\em Chandra} X-ray
  Observatory and the proposed X-ray interferometer {\em MAXIM}.

\end{abstract}


\section{Introduction}

Imaging X-ray emitting regions around black holes addresses questions
related both to black hole physics and to accretion phenomena.
Depending on the properties and the extent of the X-ray emission as
well as its proximity to the black hole, the X-ray images may have
imprinted on them the accretion geometry, its physical properties, and
even signatures of the space-time around the black hole. In standard
models, the X-ray emission from active galactic nuclei (AGN) is
assumed to arise from a hot corona above a geometrically thin
accretion disk. Thermal Comptonization of disk blackbody photons in
the corona leads to the production of a hard X-ray continuum, some of
which is reprocessed by the underlying disk (e.g., Haardt \& Maraschi
1991, 1993). From the properties of the hard X-ray continuum and of
the reflected component (e.g., Nandra et al.~1997) it is now
established that the the majority of X-ray emission from AGN
originates from regions close to the black holes, corresponding to
$\lesssim 3-20 R_{\rm{S}}$, where $R_{\rm{S}}$ is the Schwarzschild
radius of the black hole.

In recent years is has become apparent that standard thin disk
accretion with high radiative efficiency may not be common in nearby
galaxies.  The centers of most nearby galaxies also harbor
supermassive black holes (e.g., Magorrian et al. 1998) but, unlike the
more distant AGN, display little or no activity. It has been
shown that the nuclei of nearby giant elliptical galaxies, which host
the largest black holes known with masses of $10^8-10^{10} M_{\odot}$
(e.g., Magorrian et al. 1998), the center of the Milky Way with its
$2.5 \times 10^6 M_\odot$ black hole (e.g., Eckart \& Genzel 1997),
and even some of the galactic black hole sources like A0620 in
quiescence are remarkably underluminous for their expected mass
accretion rates. The relative quiescence and spectral characteristics
of these sources can be explained if the central black holes accrete
via low radiative-efficiency accretion flows or ADAFs (Rees et al.
1982; Fabian \& Rees 1995; Narayan, Yi, \& Mahadevan 1995; Narayan,
McClintock, \& Yi 1996; Di Matteo et al. 2000; for a review see, e.g.,
Narayan, Mahadevan \& Quataert 1998 and references therein).

In this class of accretion solutions the flows are two-temperature
with the electron temperature ranging from $\approx 1~\rm{MeV}$ in the
inner regions to about 1 keV in the outer regions. At these
temperatures, and when the gas is optically thin to electron
scattering so that Comptonization is unimportant, most of the cooling
occurs via thermal bremsstrahlung emission. In recent work it has been
shown that the hard X-ray components observed in elliptical galaxy
systems (Allen, Di Matteo \& Fabian 2000) are indeed consistent with
models of thermal bremsstrahlung emission (Di Matteo et al.~2000). In
the same work, it has also been emphasized that bremsstrahlung
emission may always be the dominant cooling mechanism because hot
accretion flows are likely to drive strong mass loss
(e.g. Igumenshchev \& Abramowicz 1999; Blandford \& Begelman 1999) so
that their densities and scattering optical depths always remain low
and Comptonization is never important. The same also applies to the
case in which convection becomes important in hot flows, as discussed
in recent work by Narayan, Igumenshchev \& Abramowicz (2000), Quataert
\& Gruzinov (2000), and Ball, Narayan, \& Quataert (2000).

The X-ray emission from the nuclei of nearby giant elliptical galaxies
is therefore expected to have significantly different properties than
that of AGN, reflecting the physical conditions of hot, extended
accretion flows. In this paper we go beyond the spectral calculations
carried out in earlier work and provide a set of diagnostics of
the bremsstrahlung X-ray emitting region of hot accretion flows. This
includes calculating the expected surface brightness profiles relevant
for determining the sizes and density profiles of the X-ray emitting
regions of these flows. The calculations of these quantities may have
a significant impact for explaining the ongoing and upcoming X-ray
observations of the cores of galactic nuclei. Conversely, the
measurement of these quantities may reveal the relevant mode of
accretion and the detailed properties of the accretion flows,
improving our understanding of physical processes, such as convection
and generation of outflows.  Here we discuss the implications of our
results for observations with the {\em Chandra} X-ray Observatory and
the proposed future space-based X-ray interferometer {\em MAXIM}.

In \S 2, we review bremsstrahlung emission from hot flows and
introduce the X-ray diagnostics that we study in this paper. In \S 3,
we present the surface brightness profiles and visibility curves for
different black hole masses and accretion-flow density profiles.
Finally in \S 4, we discuss the implications of our results for
current and possible upcoming X-ray missions.

\section{Bremsstrahlung Emission from Hot Flows}

\subsection{Advection Dominated Accretion}

Optically thin advection-dominated accretion flows (ADAFs) are 
examples of hot, magnetic plasmas with low radiative efficiency
(Ichimaru 1977; Rees et al. 1982; Narayan \& Yi 1994, 1995b;
Abramowicz et al. 1995; see Narayan et al.~1998 and Kato, Fukue, \&
Mineshige 1998 for reviews).  The dominant emission mechanisms in low
mass-accretion rate ADAFs are optically thin bremsstrahlung from the
hot electrons, which remain at high temperatures out to large radii in
the flow, and synchrotron radiation in the inner regions.  Above the
mass accretion rates at which the scattering optical depth approaches
unity, inverse Comptonization of soft synchrotron photons also becomes
important and a Compton peak emerges at X-ray wavelengths.  The
spectrum of a hot flow around a supermassive black hole, therefore, is
characterized by an overall low luminosity, with a synchrotron peak in
the radio wavelengths and an X-ray component coming from
bremsstrahlung emission, with possible contribution from the
Comptonization of soft synchrotron photons in the hot flow.  However
at the low mass-accretion rates which we consider here, Comptonization
is negligible and the X-ray spectra are entirely dominated by
bremsstrahlung emission.

In describing these flows, we use dimensionless quantities for the
mass, radius, and accretion rate. The black hole masses are given in
units of the solar mass, $m \equiv M / M_{\odot}$, and the radii in
units of the Schwarzschild radius, i.e., $r \equiv R / R_{\rm{S}}$,
where $R_{\rm{S}} = 2 G M / c^2$.  We scale the mass accretion rate in
units of $\dot{M}_{\rm{Edd}}$, i.e., $\dot{m} \equiv \dot{M} /
\dot{M}_{\rm{Edd}},$ where the Eddington mass accretion rate is
$\dot{M}_{\rm{Edd}} \equiv L_{\rm{Edd}} / \eta_{\rm{eff}} c^{2}$, and
$\eta_{\rm{eff}}$ is the radiative efficiency taken to be 0.1 for the
purposes of this definition. 

It has been proposed that mechanisms leading to strong outflows and
convection may be common in ADAFs. Outflows have been discussed
extensively in the literature (Narayan \& Yi 1995a; Igumenshchev \&
Abramowicz 1999; Blandford \& Begelman 1999; Stone, Pringle \&
Begelman 1999) and have been applied to the spectra of accreting
supermassive black holes (Di Matteo et al. 1999, 2000; Quataert \&
Narayan 1999). Similarly, convective processes are discussed by
Igumenshchev, Chen, \& Abramowicz (1996), Narayan, Igumenshchev \&
Abramowicz (2000), and Quataert \& Gruzinov (2000). A simple
parametrization of the accretion flow density profile in the presence
of outflows was put forth by Blandford \& Begelman (1999), in which
the mass accretion rate is modified radially and is given by $\dot{m} =
\dot{m}_{\rm out} (r/r_{\rm out})^p$, where $r_{\rm out}$ is the 
radius at which the mass accretion rate $\dot{m}_{\rm out}$ is fixed,
and $0 < p < 1$ specifies the strength of the outflow.  Other
processes such as convection may give rise to similar modifications of
the radial density profile of these flows. In particular, convective
solutions have the same density profiles as the $p=1$ inflow-outflow
solutions although the velocity profile of these solutions differ
significantly from that of inflow-outflow solutions (e.g., Narayan et
al.~2000). In this paper, we use this parametrization to calculate the
electron density profiles which scale as $N_e \propto
r^{-3/2+p}$. Throughout this paper, we fix the mass accretion rate
$\dot{m}_{\rm out}$ at $10^{-4}$ at $r_{\rm out} = 10^5$ unless
otherwise specified. The other parameters of the model describing the
microphysics are fixed at some typical values: the viscosity parameter
$\alpha = 0.1$, the ratio of gas pressure to magnetic pressure $\beta
= 10$, and the ratio of viscous electron heating to proton heating
$\delta = 10^{-2}$.

In our numerical calculations, we use the global solutions described
in Popham \& Gammie (1998) to obtain the run of electron temperature
and density with radius, with proper modifications of the continuity
equation to take into account the changes to the electron density and
temperature for $p \neq 0$. These solutions are based on those of
Narayan, Kato, \& Honma (1997) and Chen, Abramowicz, \& Lasota (1997)
but include general relativistic effects near the black hole. One
caveat in our analysis is that these global solutions are obtained for
the $p=0$ case and this may give rise to errors of order unity for
strong outflow cases (see Quataert \& Narayan 1999). To make the
interpretation of our results more intuitive, we note that the
electron temperature is nearly virial and equal to the proton
temperature in the outer regions of the flow, but the profile gets
flatter at smaller radii ($r \gtrsim 10^3$) due to synchrotron
cooling. This gives rise to the two-temperature character of the flow
at small radii. The electrons throughout the flow are assumed to be
thermal with the Maxwellian energy distribution $N_e(\gamma) = N_e
\gamma^2 \beta \exp(-\gamma/\theta) / \theta K_2(1/\theta)$, where
$\gamma$ is the electron Lorentz factor, $\theta=kT/m_e c^2$ is the
dimensionless electron temperature, and $K_2$ is a modified Bessel
function.

\subsection{Bremsstrahlung Emission}

The relativistic electron-proton bremsstrahlung emissivity is given by
(Stepney \& Guilbert 1983)
\begin{equation}
  \label{eq:bremsep}
  \frac{dE_{\rm{ep}}}{dV~dt~d\nu~d\Omega} \equiv j_{\rm{ep}} = 
 \frac{h c N_p}{4 \pi}  \int_{1+\omega}^\infty 
\omega \frac{d\sigma}{d\omega} \beta N_e(\gamma) d\gamma
\end{equation}
where for the differential cross-section $d\sigma/d\omega$ we have
used the Bethe-Heitler formula derived in the Born approximation
(Jauch \& Rohrlich 1959).  Here, $\omega \equiv h \nu / m_e c^2$ is the
dimensionless photon energy, $N_p$ is the proton number density and
$N_e(\gamma)$ is the electron density given above. Stepney \& Guilbert
(1983) provide fitting functions for the electron-electron
bremsstrahlung emissivity given by
\begin{equation}
  \label{eq:bremsee}
j_{\rm{ee}} = N_e^2 \sigma_{\rm{T}} \alpha_{\rm{f}}
 m_e c^3 \exp(-x) G(x,\theta)/4 \pi x
\end{equation}
where $\sigma_{\rm{T}}$ is the Thomson cross-section,
$\alpha_{\rm{f}}$ is the fine-structure constant, $x \equiv
\omega/\theta$, and the fitting function $G(x,\theta)$ can be found in
tabular form in Stepney \& Guilbert (1983). Typically, $j_{\rm{ee}}
\lesssim 0.1 j_{\rm{ep}}$ for the frequencies and temperatures
considered here.

To calculate the emerging bremsstrahlung intensities, we use a
one-dimensional radiative transfer algorithm as described by \"Ozel,
Psaltis, \& Narayan (2000). We integrate the equation of radiative
transfer along plane-parallel rays of varying impact parameters $b$
(perpendicular distances of rays to the central line of sight) through
the flow and calculate the emerging intensity (i.e., surface
brightness profile) $I(b)$. Note that in our radiative transfer
calculations we do not take into account general relativistic redshift
and light deflection which will be important for centrally peaked
profiles $(p \sim 0)$ and may lead to large errors when most of the
emission comes from within a few $R_S$ of the event horizon. Our
numerical domain extends to an impact parameter of $\sim 10^6
R_{\rm{S}}$, beyond which $kT_e \ll 1$~keV and bremsstrahlung
radiation does not contribute to the X-ray energies.  Also note that
our treatment assumes spherical symmetry of the density and
temperature profile of the electrons which is a reasonable assumption
given that all the flows considered here have large scale heights,
with $H/R \approx 1$. For optically thin bremsstrahlung emission in
such extended flows, the region around the event horizon, i.e., small
radii, have small contribution to the emerging intensity, even at
small impact parameters.

\subsection{X-ray Diagnostics}

The X-ray spectra of hot flows around supermassive black holes have
been studied extensively in previous work. Here we consider additional
powerful diagnostic quantities that are related to the two primary
imaging methods: the brightness profiles of accretion flows, which are
directly measurable by single-aperture imaging in X-rays, and the
Michelson visibility function, which is the curve measured by an
interferometer. Note that two of these diagnostic methods, spectral
measurements and CCD imaging, are being carried out with increasing
resolution by current X-ray telescopes. Using interferometry to study
the sizes and profiles of accretion flows is being planned for future
X-ray missions.

The brightness profile $I(b)$ of a spherically symmetric source is the
map of the line-integral of the emissivity as a function impact
parameter $b$. To define the visibility function, we first introduce
the projected brightness profile $B(x)$. This is defined as the
integral of specific intensity in the direction orthogonal to the
interferometer baseline; i.e., the aperture spacing vector $x$ (see,
e.g., Thompson, Moran, \& Swenson 1991) and is given by
\begin{equation}
\label{eq:pbt}
B(x) \equiv \int I(\sqrt{x^2+y^2}) dy. 
\end{equation}
Thus $B(x)$ is effectively a surface-integral of the emissivity. The
Michelson visibility function is then
\begin{equation}
  \label{eq:fft} 
V_M(k) \equiv \frac{\vert \int B(x) \exp(ikx) dx  \vert}
{\vert \int B(x) dx \vert}
\end{equation}
which is the normalized modulus of the Fourier transform of the
projected brightness profile $B(x)$. The visibility function thus
probes the characteristic length scales of the flow and can be
inverted to give the brightness profile of the source. Finally, the
luminosity at each photon frequency for optically thin emission is the
volume integral of the emissivity and is a measure of the total energy
output from the entire flow at that frequency.

Given the above definitions, we point out that, for optically thin
bremsstrahlung emission, the brightness profile, the projected
brightness profile, and the luminosity are effectively (but not
strictly) the three radial moments of bremsstrahlung emissivity $j$.
Because of this, as well as due to the strong dependence of
bremsstrahlung emissivity $j$ on electron density ($j \propto n^2$),
intensity $I_\nu$ and projected brigtness profile $B_\nu$ are strong
indicators of the density profiles of the hot flows.

\section{Results: Intensity and Visibility}

Figure~1 shows the emerging intensity at 1 and 10~keV as a function of
impact parameter for a hot accretion flow around a $10^9 M_\odot$
black hole. The solid curves show the intensity profile at 1~keV while
the dashed curves correspond to 10~keV.  As expected from the
dependence of intensity on the density profile, flows with strong
winds or convection (p=1) have flat radial intensity profiles
extending up to the radii at which the electron temperature drops
below 1 and 10~keV, corresponding to $r \sim 10^6$ and $r \sim 10^5$
respectively. The profiles steepen significantly as the parameter $p$
decreases, with the steepest profile corresponding to the case with no
outflows. The 10~keV emission is suppressed overall with respect to
the 1~keV emission and cuts off exponentially at $\approx 0.1 $ of the
radius of the 1~keV emitting region.  This is because the temperature
is nearly virial at these radii and order of magnitude difference in
temperature corresponds to roughly the same difference in
$r$. Finally, we note that since the hot flows show self-similar
behaviour, the intensity $\it{profile}$ does not change with changing
black hole mass, even though the normalization of the intensity scales
with mass.

The {\it extent} and {\it size} of hot accretion flows show
significantly different responses to changing outflow strength. By
size, we mean the full-width half maximum (FWHM) of the visibility
curve, while by extent we refer to the radius marking the exponential
cutoff of emissivity (full-width zero intensity). The full-width zero
intensity is very insensitive to $p$; irrespective of the steepness of
the density profiles, all flows extend out to large radii before the
exponential cut-off.  However, the steepness of the brightness
profiles changes dramatically, with most of the emission arising from
within $\sim 6 R_S$ for $p=0$ and within $10^4 R_S$ for $p=1$
(Figure~1).  Therefore, the corresponding sizes also change by orders
of magnitude (see Figure~2). 

Figure~2 shows the Michelson visibility curves as a function of
angular wavenumber $k_\theta = 1 / \theta$, where $\theta$ is the
angular size of the X-ray region in arcseconds. The top axis shows the
corresponding baseline needed to resolve a scale $k_\theta$ at 1~keV.
The dashed lines are the visibility functions for a 10 $M_\odot$
galactic black hole at 3~kpc distance, while the solid curves show the
case for a $10^9 M_\odot$ black hole at 18~Mpc (Virgo cluster)
distance. Both of these are plotted for different outflow strengths,
$p=0, 0.5$, and 1. Finally, as a specific example, we plot with the
dotted line the visibility function for a hot accretion model of M87,
with a density profile that best fits the observed spectrum of this
source (with $\dot{m}_{\rm out} = 0.015$, $r_{\rm out} = 300$, $p =
0.45$, Di Matteo et al.~2000). 

\vspace*{0.3cm}
\vbox{ \centerline{ \psfig{file=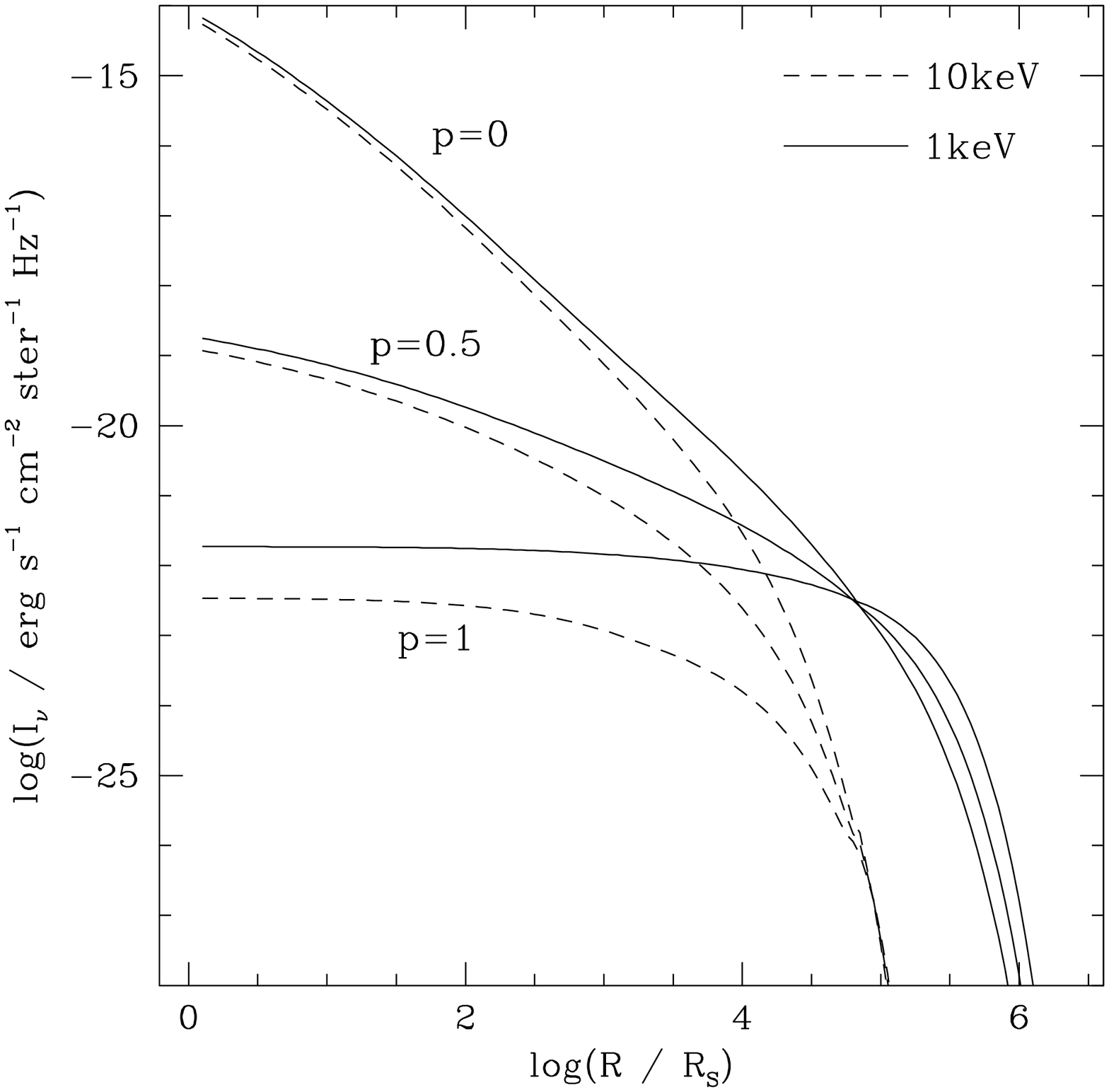,width=9.0truecm} }
\figcaption[]{ \footnotesize The surface brightness as a function of
  impact parameter in units of Schwarzschild radius $R_{\rm{S}}$ for
  three values of the outflow parameter $p=0$, $0.5$, and 1 for a
  black hole mass $m=10^9$. The solid lines shows the bremsstrahlung
  intensities at 1~keV, while the dashed lines correspond to 10~keV.
  }}
\vspace*{0.5cm}

The typical observed angular sizes of these flows range from $\sim
0.01-1$ arcsec for the case of the supermassive black hole and from
$1-100$ $\mu$arcsec for the galactic black hole.  This is in
accordance with the profiles shown in Figure~1, which shows that the
characteristic angular size of the bremsstrahlung emitting region
increases significantly in the presence of an outflow. Specifically,
for strongly modified density profiles ($p=1$), the flow sizes are
large, typically corresponding nearly to their whole extent, $10^6
R_{\rm{S}}$.  This is because the intensity profile is flat and
therefore the only characteristic scale in the flow is the size of the
entire 1~keV emitting region. On the other hand, for intensities that
are radially peaked, i.e.,when $p \lesssim 0.5$, the power is
distributed over a broader range of scales and the characteristic size
of the flow is smaller.  Specifically, a purely inflow ADAF is
extended but still centrally peaked, with a measurable FWHM of only
tens 

\vbox{ \centerline{ \psfig{file=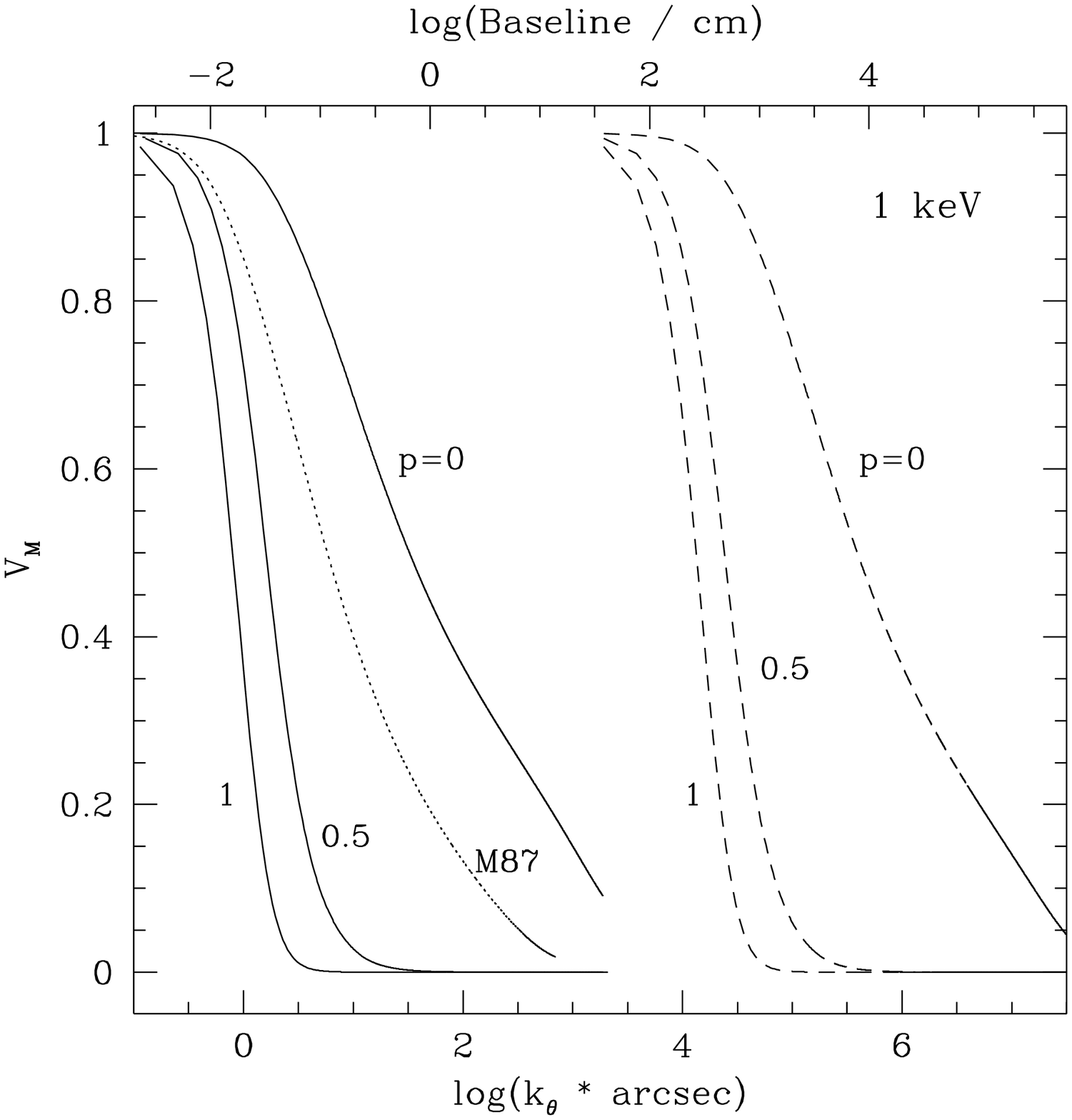,width=9.0truecm} }
\figcaption[]{ \footnotesize The solid lines show the Michelson
  visibility curves for a $10^9 M_\odot$ black hole at a distance of
  the Virgo cluster (18 Mpc) for p=0, 0.5, and 1. The dashed curves
  correspond to a $10 M_\odot$ black hole at 3kpc distance. The dotted
  line is the visibility function for a specific model of M87 with a
  density profile that best fits the observed spectrum of this source
  (Di Matteo et al. 2000). The bottom axis shows the angular
  wavenumber in units of inverse arcseconds while the top axis shows
  the baseline that would be required for resolving a wavenumber
  $k_\theta$ at 1keV. }}
\vspace*{0.5cm}

\noindent of $R_{\rm{S}}$ (for the $p=0$ case, ${\rm FWHM} = 30$).

The complementary nature of the two imaging methods, namely imaging
which probes directly the intensity profiles, and interferometry which
measures the visibility function, is evident in Figures~1 and 2. The
extended flows around supermassive black holes with flat profiles have
large characteristic angular sizes extending up to 1 arcsec, which may
be resolved with direct imaging instruments (see \S 4 for a discussion
of specific instruments). On the other hand, an interferometer with a
baseline of $\gtrsim 1m$ is not sensitive to most of the power on
large scales and is thus not suitable to study these flows.
Furthermore, since the flat density profile flows are also less
luminous, it is advantageous to observe these sources with direct
imaging instruments which, unlike an interferometer, allow for a
photon-energy integrated signal. The situation, however, is reversed
for flows with radially peaked intensity profiles.  Here, a good
resolution that is necessary to study the characteristic sizes of a
few tens of $R_{\rm{S}}$ can be achieved only by an interferometer.
Therefore, both interferometric and direct imaging studies are needed
to study the varying characteristics of hot accretion flows. In \S4,
we discuss further the resolution requirements for upcoming
instruments which would allow the current direct imaging instruments
and interferometry to work in this complementary manner.

\vbox{ \centerline{ \psfig{file=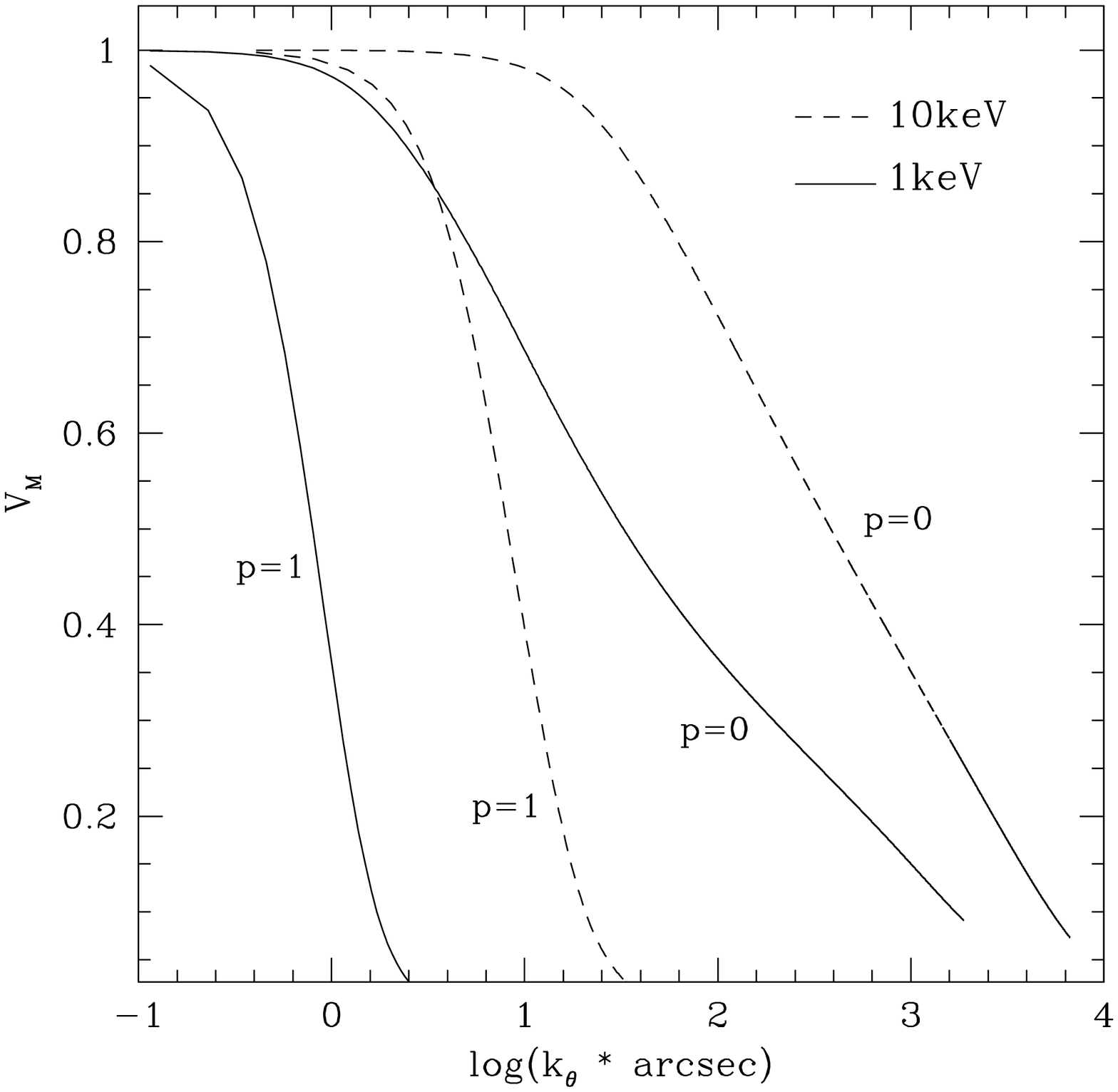,width=9.0truecm} }
\figcaption[]{ \footnotesize The change in the visibility functions
  from 1~keV to 10~keV for a black hole mass $m=10^9$ and p=0 and 1.
  Both the sizes of the flow and the visibility functions depend
  strongly on photon energy. }}
\vspace*{0.5cm}

In Figure~3, we show the change in the visibility function from 1~keV
to 10~keV. The characteristic size of the 10~keV emitting region
decreases by an order of magnitude for both $p=1$ and $p=0$.
Furthermore, since the 10~keV brightness profile falls off more
steeply in radius for $p < 1$ (Fig.1), the visibility functions are
broader.

\section{Discussion}

High resolution X-ray imaging and interferometric studies, which
measure directly or indirectly the brightness profiles and thus the
sizes and density profiles of accretion flows around black holes, can
be powerful probes for studying the relevant mode of accretion and the
detailed properties of accretion flows. We have shown that the 1~keV
emitting regions of hot accretion flows can be large and extend up to
$10^6 R_{\rm{S}}$ irrespective of the specifics of the density
profile, provided that the ADAF itself extends out to this radius (see
below for further discussion). However, the brightness profiles and
visibility functions are strongly affected by the specific density
profile. In particular, the flat density profiles in the presence of
strong outflows (p=1) give rise to very flat surface brightness
profiles with characteristic sizes of $10^6 R_{\rm{S}}$. In contrast,
the purely inflow advection dominated solutions have centrally peaked
profiles with characteristic sizes of only a few tens of $R_{\rm{S}}$.

\vbox{ \centerline{ \psfig{file=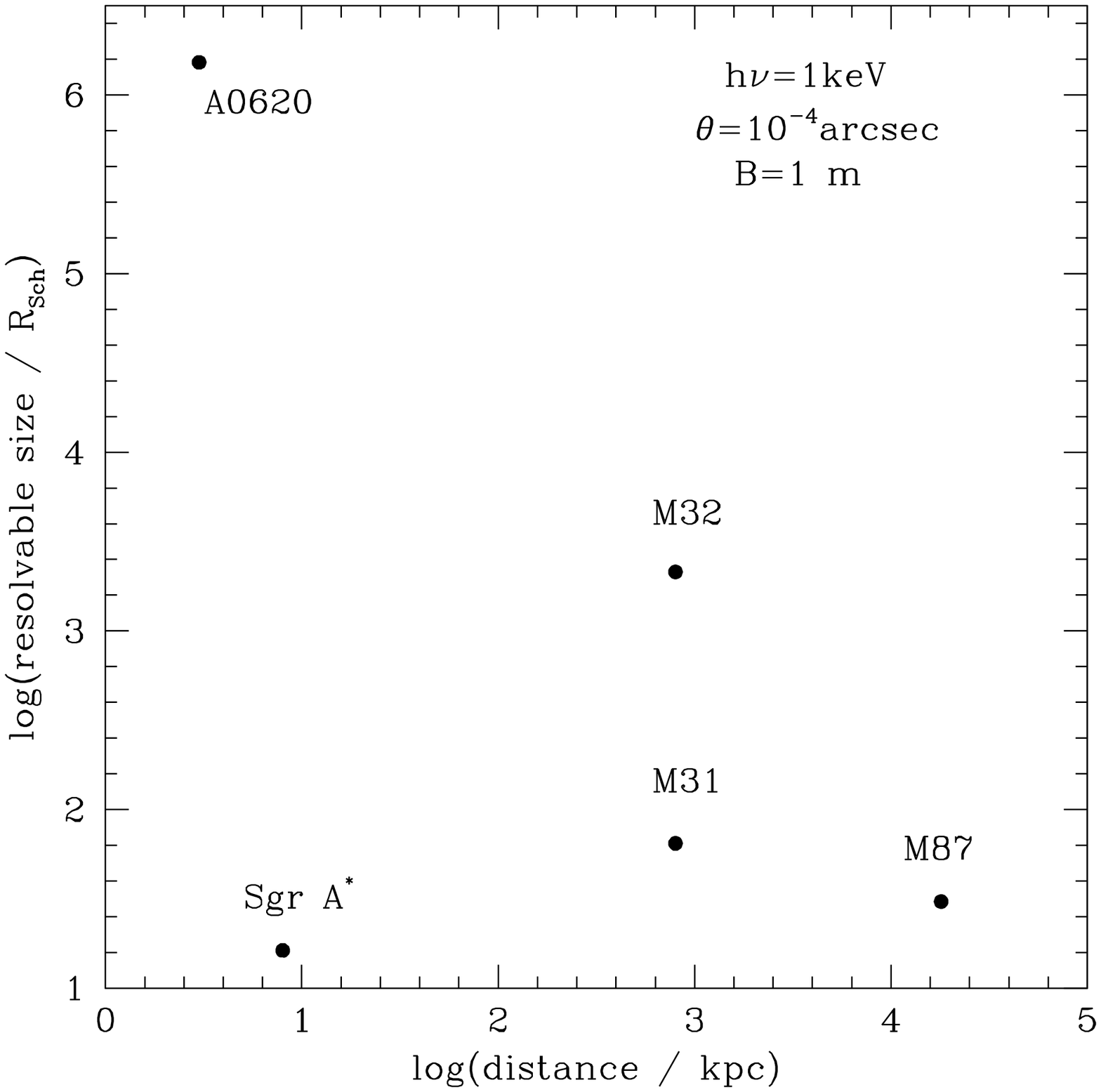,width=9.0truecm} }
\figcaption[]{ \footnotesize The resolvable scale in units of
  $R_{\rm{S}}$ of each source plotted against the known distance to
  the sources.  If the black holes are accreting via a hot flow, these
  resolvable sizes represent the largest size the accretion flow can
  have to be resolved by a $\it{minimum}$ baseline of 1m. }}
\vspace*{0.5cm}

If we are correct in surmising that hot accretion flows may be common
in nearby galactic nuclei and that such flows may drive strong
outflows (e.g., Quataert \& Narayan 1999; Di Matteo et al. 2000) then
the above results have strong implications for X-ray observations of
these nuclei. An important consequence of our results is that a hot
accretion flow with a flat density profile around a $\sim 10^9
M_\odot$ black hole at Virgo cluster distances, such as M87, may look
like an extended X-ray source rather than a point source when imaged
by the $\it{Chandra}$ X-ray observatory with its 0.5 arcsec resolution
mirrors. At the distance of M87 and for $m \sim 3\times 10^9$
(Macchetto et al. 1997), 0.5 arcsec corresponds to $10^5 R_{\rm{S}}$
which is about an order of magnitude smaller than the extent of the
regions of these flows emitting soft X-rays (at $\lesssim 2$ keV;
where most of the photon counts would be at these fluxes). Conversely,
the detection of an X-ray point source in elliptical nuclei in Virgo
must imply that either the physical sizes of the hot accretion flows
are smaller than $10^5 R_{\rm{S}}$ or that standard thin disk
accretion with its associated corona applies. Regarding the former
possibility, we stress that in our calculation we have assumed that
the hot flows extend to their maximum possible X-ray emitting radius,
i.e., $R \sim 10^6 R_{\rm{S}}$. This radius is consistent with the
typical accretion radius (the radius at which the black hole
gravitational potential dominates that of the galaxy) of elliptical
galaxies ($\sim 0.1$ kpc; Di Matteo et al. 2000) but does not need to
coincide with the radius at which the angular momentum dominated
accretion flow, i.e. an ADAF, is formed. There may be physical
processes such as cooling at large radii that may cause the ADAF flow
to start at smaller radii, or there may be transitions to other
accretion solutions in the outer regions (see, e.g., Igumenshchev,
Abramowicz, \& Novikov 1998; Menou et al. 1999). Here, we assumed a
direct transition from the hot interstellar medium (ISM) of the
elliptical galaxies to an ADAF.

We note also that this hot ISM of the elliptical galaxies, with
typical temperature of a few keV, may also contribute significantly to
the emission in the soft X-ray band, which would further blend the
images and may make the detection of the accretion flow component more
difficult. However, in the hard band ($> 5$~keV), the sources are
intrinsically smaller (c.f., Fig.~3), and may look more point-like
provided that there are enough photon counts. This would have the
additional advantage of having less background from galactic emission.
Finally, we stress that {\em Chandra} X-ray spectroscopy of the
central regions of these galaxies will also complement the information
provided by the imaging instruments and will be crucial for
determining the presence of ADAFs in these systems. With its 1 MeV
temperature, the bremsstrahlung emission from the central regions of a
hot accretion flow will be much harder than that of the galaxy ISM and
will have a different spectral slope than that expected from a typical
AGN with low luminosity.

Variability studies are also important for determining the sizes of
accretion flows and are complementary to imaging studies. Rapid
variation in the X-ray flux may help constrain the size of the X-ray
emitting region. For example, {\it ROSAT} HRI observations (at $E \sim
1$ keV) of the ``core'' of M87 show $\approx 20 \%$ variability on
timescales of $\approx 6$ months to a year (Harris, Biretta, \& Junor
1997).  This is not easy to reconcile with a $p > 0.5$ bremsstrahlung
model for the X-ray emission. However, identifying the possible
contributions to the variability from jets associated with the
accretion flows relies on high resolution imaging studies and
therefore the two methods can be used as complementary probes.

Because of the large inferred black hole masses, many of the nearby
supermassive black holes have also been considered some of the best
targets for observations with the planned X-ray interferometry mission
{\em MAXIM}.  Owing to the short X-ray wavelengths, remarkably high
spatial resolutions can be achieved with relatively short baselines
with the ultimate aim of imaging of black holes on even horizon
scales. (See http://maxim.gsfc.nasa.gov/ for a detailed discussion of
the proposed capabilities of {\em MAXIM} and Cash et
al.~2000). Already for the first stages of such a mission, a target
resolution of $0.1$ mas has been planned corresponding to a
baseline of $\sim 1~m$ at 1~keV. This should probe up to a few tens
of $R_{\rm{S}}$ for the supermassive black holes in nearby galaxies
(Fig~4).

However, if indeed large black holes in nearby galaxies accrete via
hot, radiatively inefficient accretion flows, e.g, M87, one important
consequence of our results is that characteristic X-ray emission scale
would be large and there would be little power on scales corresponding
to few tens of $R_{\rm{S}}$ or smaller. Therefore baselines of $\sim
1$~m may not be appropriate to image black holes in the Virgo
ellipticals (c.f., \S 3 and Fig.~2) and a range of smaller baselines
would be necessary to study the visibility functions of these extended
flows. Thus, to study properties of accretion flows, the minimum
baseline of the interferometer is an important consideration.
Specifically, for studying the spatial extent of hot accretion flows
and the relevance of outflows, it is important to bridge a possible
gap in resolution between that of the $\it{Chandra}$ X-ray observatory
and the minimum resolution of an X-ray interferometer.  Alternatively,
with a baseline of $\sim 1$~m, galactic black hole candidates may be
better targets to study hot accretion flows (Fig.~2). $\it{Chandra}$
observations will be crucial for selecting the appropriate targets for
an X-ray interferometry mission and giving us clues on the relevant
mode of accretion in the nearby black holes.

\acknowledgements We thank Ramesh Narayan for many useful suggestions
throughout this work. We also thank Dimitrios Psaltis, Martin White,
Irwin Shapiro, and George Rybicki for valuable discussions and
comments on the manuscript.  T.\,D.\,M.\ acknowledges support for this
work provided by NASA through Chandra Postdoctoral Fellowship grant
number PF8-10005 awarded by the Chandra Science Center, which is
operated by the Smithsonian Astrophysical Observatory for NASA under
contract NAS8-39073. This work was supported in part by NSF Grant AST
9820686.

\end{document}